\documentclass[]{tPHM2e}

\begin{document}
\doi{10.1080/14786435.20xx.xxxxxx}
\issn{1478-6443}
\issnp{1478-6435}
\jvol{00} \jnum{00} \jyear{2014}
\markboth{Taylor \& Francis and I.T. Consultant}{Philosophical Magazine}

\title{
Ultrasonic Study of the Hidden Order and Heavy-Fermion State in URu$_2$Si$_2$ with Hydrostatic Pressure, Rh-doping, and High-Magnetic Fields
}

\author{Tatsuya Yanagisawa$^{\ast}$\thanks{$^\ast$This is an Author's Original Manuscript of an article submitted for consideration in the Philosophical Magazine (\copyright  Taylor\&Francis); Philosophical Magazine is available online at http://www.tandfonline.com/loi/tphm20\vspace{6pt}}\\\vspace{6pt} 
{\em{
Division of Physics, Faculty of Science, Hokkaido University, Sapporo 060-0810, Japan
}}{\em{}}\\\vspace
{6pt}\received{27 Oct. 2013} }
\maketitle

\begin{abstract}

This paper reports recent progress of ultrasonic measurements on URu$_2$Si$_2$, including ultrasonic measurements under hydrostatic pressure, in pulsed-magnetic fields, and the effect of Rh-substitution. The observed changes of the elastic responses shed light on the orthorhombic-lattice instability with $\Gamma_3$-symmetry existing within the hidden order and the hybridized 5$f$-electron states of URu$_2$Si$_2$. 

\bigskip
\end{abstract}

\section{Introduction}

Ultrasonic measurement is a powerful tool to probe electric multipole responses and symmetry-breaking lattice instabilities in single crystals; using the bulk elastic response of solid states via electron-phonon coupling, by means of sound-velocity measurement, it produces accuracies up to $\Delta v/v \sim 10^{-8}$. Initially, ultrasonic measurements on URu$_2$Si$_2$ were reported individually by German and Japanese groups in the 1990fs \cite{1,2,3,4,5,6}; there have also been other papers regarding ultrasonics in the superconductivity ($T_{\rm c} \sim 1.4 $K) of this compound and magnetic field dependence of elastic constants in the last decade \cite{7,8,9,10,11}. Figure 1(a) represents the normalized elastic constants of URu$_2$Si$_2$ as a function of temperature, recently re-measured by our group. It can clearly be seen that the longitudinal $C_{11}$ and transverse $(C_{11}-C_{12})/2$ modes show elastic softening; a decreasing of the elastic constant with decreasing temperatures, from $\sim$ 100 K and $\sim$120 K, respectively, while the other two shear modes $C_{44}$ and $C_{66}$ are increasing monotonically. These results are consistent with the previous reports \cite{12,13,14}.

\begin{figure}
\begin{center}
\resizebox*{12cm}{!}{\includegraphics{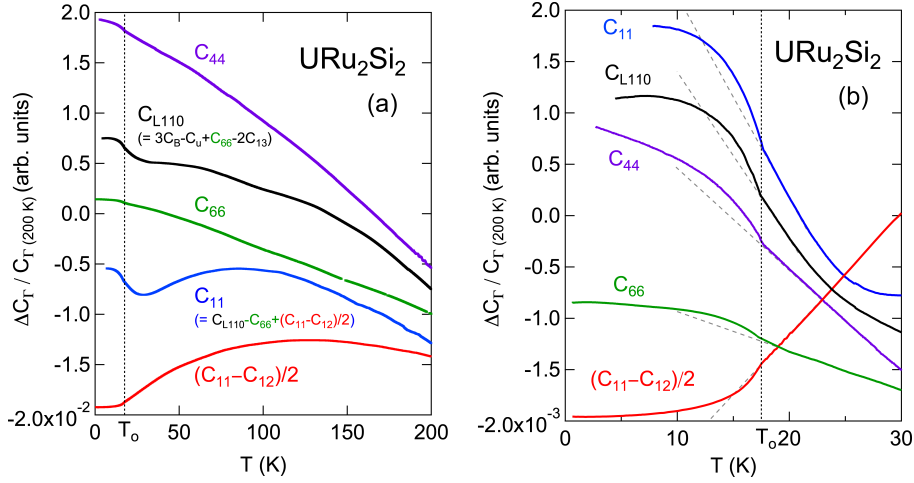}}
\caption{(a) Relative change of elastic constants in URu$_2$Si$_2$ as a function of temperature. (b) Enlarged plot for low temperature region around the hidden order.}%
\label{fig01}
\end{center}
\end{figure}

The most characteristic feature of the elastic constants around the hidden order (HO) transition ($T_{\rm o}$ = 17.5 K) \cite{15, 16, 17} is that $C_{11}$ and $C_{\rm L[110]}$ exhibit local minima at around 20 K and upturns toward the HO transition. In the early stages of ultrasonic studies on this compound, this characteristic temperature dependence of $C_{11}$ had been analyzed by considering strong Gr{\"u}neisen coupling, caused by a many-body effect; in other words, a `volume collapse', which is generally found in heavy-fermion compounds \cite{18, 19, 20}. However, after finding clear softening in the shear mode $(C_{11}-C_{12})/2$, which corresponds to the $\Gamma_3$-symmetry-breaking strain \cite{5}, the Gr{\"u}neisen coupling scenario should be reconsidered, because the volume collapse only affects the bulk-modulus, {\it i.e.}, it is related to a volume change with $\Gamma_1$ total symmetry.

In Table 1, the symmetry, strain, and related elastic-stuffiness constants with their absolute values are summarized for URu$_2$Si$_2$ at 4.2 K. For tetragonal symmetry, we can rewrite the $C_{11}$ as a sum of the elastic constants with symmetrized representations; $C_{11} = 3C_{\rm B}-C_{\rm u}+(C_{11}-C_{12})/2-2C_{13}$, where $C_{\rm B} = (2C_{11}+C_{12}+4C_{13}+C_{33})/9$ is the bulk modulus, $C_{\rm u} = (C_{11}+C_{12}-4C_{13}+2C_{33})/6$ is a tetragonal mode \cite{21}, which is related to the tetragonal strain $\epsilon_{\rm u}=(2\epsilon_{zz}-\epsilon_{xx}-\epsilon_{yy})/\sqrt{3}$, and belong to $\Gamma_1$ symmetry in the tetragonal symmetry (see Table 1). Now it is apparent that $C_{11}$ includes not only the bulk modulus but also the component of $(C_{11}-C_{12})/2$ with $\Gamma_3$ symmetry. In order to check the contribution of the bulk-modulus change in this compound, the longitudinal ultrasound propagated along [110] axis, which corresponds to $C_{\rm L[110]}$, was recently measured. Here, the ultrasonic constant $C_{\rm L[110]}$ is rewritten as $C_{\rm L[110]} = 3C_{\rm B}-C_{\rm u}+C_{66}-2C_{13}$, which also includes the bulk modulus but also includes $C_{66}$ instead of $(C_{11}-C_{12})/2$.  The $C_{66}$ does not show any anomalies except for a tiny slope change at the HO transition, so comparing this to $C_{\rm L[110]}$, we can estimate that the softening of $C_{11}$ from 100  to 20 K mostly originates in the softening of $(C_{11}-C_{12})/2$. Indeed, the upturn-like bulk modulus change is dominant below 20 K; thus, the Gr{\"u}neisen-coupling effect on $C_{\rm B}$ does exist in the temperature region below 20 K.

\begin{table}
\tbl{Symmetry, strain, multipole and related elastic stuffiness constant with absolute value}
{\begin{tabular}{@{}cccc}\toprule
Symmetry			&Strain			&Elastic Stiffines Constant			&Absolute Value$^{\dag}$
\\
\colrule
$\Gamma_{\rm 1g}$ (A$_{\rm 1g})$ 	& $\epsilon_{\rm B}=\epsilon_{xx}-\epsilon_{yy}-\epsilon_{zz}$ 			& $C_{\rm B} = (2C_{11}+C_{12}+4C_{13}+C_{33})/9$	&-		\\
$\Gamma_{\rm 1g}$ (A$_{\rm 1g})$ 	& $\epsilon_{\rm u}=(2\epsilon_{zz}-\epsilon_{xx}-\epsilon_{yy})/\sqrt{3}$	& $C_{\rm u} = (C_{11}+C_{12}-4C_{13}+2C_{33})/6$	&-		\\
$\Gamma_{\rm 1g}$ (A$_{\rm 1g})$	& $\epsilon_{zz}=\epsilon_{\rm B}/3-\epsilon_{\rm u}/\sqrt{3}$ 			& $C_{33} = -3C_{\rm B}+4C_{\rm u}+C_{13}$					&-		\\
$\Gamma_{\rm 3g}$ (B$_{\rm 1g})$ 	& $\epsilon_{xx}-\epsilon_{yy}$										& $(C_{11}-C_{12})/2$						&6.45	\\
$\Gamma_{\rm 4g}$ (B$_{\rm 2g})$ 	& $\epsilon_{xy}$												& $C_{66}$									&12.6	\\
$\Gamma_{\rm 5g}$ (E$_{\rm g})$ 	& $\epsilon_{yz}, \epsilon_{zx}$										& $C_{44}$								&9.72	\\
\colrule
$\Gamma_{\rm 1g} \bigoplus \Gamma_{\rm 3g}$ & $\epsilon_{xx}, \epsilon_{yy}$	 						& $C_{11} = 3C_{\rm B}-C_{\rm u}+(C_{11}-C_{12})/2-2C_{13}$	&26.8	\\
$\Gamma_{\rm 1g} \bigoplus \Gamma_{\rm 4g}$ & $\epsilon_{\rm 110}=\epsilon_{\rm B}/3-2\epsilon_{\rm u}/\sqrt{3}+\epsilon_{xy}$ 	& $C_{\rm L[110]} = 3C_{\rm B}-C_{\rm u}+C_{66}-2C_{13}$	&35.1\\
\botrule
\end{tabular}}
\tabnote{$^{\dag}$Measured at 4.2 K. The unit is $10^{10}$  J m$^{-3}$.} 
\label{symbols}
\end{table}

At the HO transition, all of the elastic constants show small but clear anomalies (slight changes of slope), as we can see in Fig. 1 (b). The slight shift of the normalized elastic constants below $T_{\rm o}$ are quite small, and are roughly estimated as 0.03\% for $C_{44}$ and 0.002\% for $C_{66}$. These values are varied, but the order of the magnitude is same as in previously reported papers from different groups \cite{3, 5, 14}. Therefore, these small variations could be caused by sample quality and/or the polishing conditions of the sample surfaces, or even possibly different measurement frequencies, due to the change of `directivity' of the sound wave. In any case, the magnitude of these elastic anomalies is still two orders of magnitude smaller than the anomaly at the antiferro-quadrupolar (AFQ) order in the typical AFQ compounds of $4f$-electron systems \cite{22,23}.
Note that there are contradictions concerning the small elastic anomaly at $T_{\rm o}$ and the theoretical explanation of recent magnetic-torque measurements, which have been interpreted as spontaneous 4-fold rotational-symmetry breaking in the tetragonal basal plane that takes place in the HO phase\cite{24}. In order to explain the torque-measurement results, Thalmeier {\it et al.} and Ikeda {\it et al.} have proposed $\Gamma_5^+$ (E$^+$)-type quadrupole and $\Gamma_5^-$ (E$^-$)-type dotriacontapole order parameters, respectively (here the sign of + and - indicate the parity of the time-reversal symmetry) \cite{25, 26}. If these theoretically proposed symmetry breakings of the electronic system cause ($yz$, $zx$)-type ($\Gamma_5$) or $xy$-type ($\Gamma_4$) tiny local-lattice distortions via finite electron-phonon interactions, the elastic constants $C_{44}$ or $C_{66}$ are expected to show  relatively large anomalies at $T_{\rm o}$, respectively. At present, no distinct softening above $T_{\rm o}$ and change of $C_{44}$ and $C_{66}$ has been observed at $T_{\rm o}$ except for a tiny kink at $T_{\rm o}$, which could be caused by a thermal expansion effect of $\Delta L/L \sim 10^{-5}$ in the sound velocity measurement \cite{de Visser}. In addition, no evidence is provided for structural-symmetry breaking in URu$_2$Si$_2$ from microscopic measurements such as $^{29}$Si-NMR or high-precision x-ray and neutron scattering, thus far \cite{Takagi, Amitsuka, Tabata}. Therefore, the above multipole-order models require some assumptions that a multipole-strain interaction is extremely small or that a higher-coupling exists, {\it e.g.}, quadratic strain coupling \cite{25}, to understand the tiny elastic anomalies at $T_{\rm o}$.
Then, we return to one simple question; what is the origin of the distinct softening of $(C_{11}-C_{12})/2$? In order to discuss the origin of the softening above $T_{\rm o}$, an appropriate background of the temperature dependence of the elastic constant must be estimated because the temperature dependence of this elastic constant includes acoustic phonon contributions due to anharmonic oscillations of the lattice as well as the response of the electron systems \cite{27}. Comparing with the elastic properties of ThRu$_2$Si$_2$, where the elastic anomalies are absenct \cite{13}, we can at least conclude that the softening of URu$_2$Si$_2$ originates with the $5f$-electrons of U.
Here, based on the localized f-electron picture of the crystalline electric field (CEF) ground state, the temperature and magnetic field dependences of the elastic constant $(C_{11}-C_{12})/2$ are generally understood as a susceptibility of $\Gamma_3$-type charge distributions, such as a quadrupole moment of $O_2^2$ (= $J_x^2-J_y^2$) in the tetragonal symmetry. On the other hand, the low-temperature $5f$-electronic state of URu$_2$Si$_2$ is, however, more likely to have itinerant character due to strong hybridization. Since there is also the important fact that the $(C_{11}-C_{12})/2$ mode is the only ultrasonic mode which exhibits softening in this compound and is also related to the symmetry-breaking strain field, this anisotropic and mode-selective elastic response of URu$_2$Si$_2$ is reminiscent of the symmetrical potential deformation \cite{27, 28, 29} due to the Jahn-Teller effect of the $c$-$f$ hybridized band, the so called `Band Jahn-Teller (BJT) effect', rather than the CEF model with (pseudo-) degenerate ground states, which potentially also causes elastic softening in other transverse modes. The problem is that such a BJT-type potential deformation generally accompanies a structural change of the lattice \cite{Iron-pnictides}, as is confirmed by band-structure calculations and microscopic-structural analysis. Since URu$_2$Si$_2$ keeps tetragonal symmetry even in the HO phase as described above, we expect that a putative band deformation would be of a staggered type, instead of the uniform type observed in the BJT compounds. Thus, we should not simply apply the standard BJT scenario for the present compound, and some exotic effects such as higher-order coupling mechanisms may need to be considered to interpret the present results.
In the following chapters, we will focus on the elastic constants $(C_{11}-C_{12})/2$ and $C_{11}$ which associate with $\Gamma_3$-symmetry, and investigate the relation between the lattice instability, the HO and the heavy fermion state of URu$_2$Si$_2$ through ultrasonic measurements in hydrostatic pressure (Chap. 2), with Rh-doping (Chap. 3), and in pulsed-magnetic fields (Chap. 4).

\section{Ultrasonics under Hydrostatic Pressure}

In this chapter, we report success in the first ultrasonic measurements of URu$_2$Si$_2$ under hydrostatic pressures. The HO phase of URu$_2$Si$_2$ transitions to a type-I large-moment antiferromagnetic (LMAF) phase, with a moment of a 0.4$\mu_{\rm B}$ and a propagation vector of $Q$ = (001), which is separated from the HO by a first-order phase transition at $\sim$ 0.5 GPa \cite{30}. Figure 2 (b) represents the temperature-pressure ($T$-$P$) phase diagram for URu$_2$Si$_2$. Notably, there are reports that the tiny antiferromagnetic (SMAF) moment (0.01-0.03$\mu_{\rm B}$, depending on sample quality) observed in the HO phase is induced by strain that causes a small amount of the neighboring LMAF phase to coexist with the HO phase even in ambient pressure \cite{31,32}. Here, it is expected that the induced SMAF moment can easily modulate not only the measurement of magnetic properties but also lattice properties, due to magneto-elastic coupling. We, therefore, performed ultrasonic measurements of URu$_2$Si$_2$ under hydrostatic pressures to check the effect of the pressure-induced LMAF and SMAF on the elastic constants.
To apply pressure, we use a Cu-Be and Ni-Cr-Al hybrid piston pressure cell with an ultra-thin semi-rigid coaxial cable, which is carefully inserted into the high-pressure space. Daphne oil 7373 was used as pressure-transmitting medium in the present measurements. Figure 2 (a) shows the elastic constant $C_{11}$ as a function of temperature, measured with a frequency of 103 MHz by using an 180 deg. hybrid junction, under pressures up to 1.6 GPa and temperatures from 2  to 300 K. The data are shifted vertically with an equally spaced offset for clarity. The $C_{11}$ exhibits a sudden decrease in the vicinity of the pressure-induced LMAF transition $T_{\rm M}$, which is different from the elastic anomaly at the HO, where $C_{11}$ increases with a change of curvature at $T_{\rm o}$. Interestingly, the elastic anomaly at the HO transition under 0.95 GPa also shows tiny step-like decreases of 0.004\% compared to relatively large decreases of 0.15\% at the LMAF transition. The change of the elastic response at the HO transition at $\sim$0.95 GPa implies the possible effect of the SMAF moment, which exists parasitically in the HO phase under pressure. The temperature where $C_{11}$ shows a broad local maximum and minimum above $\sim$30 K, shifts to higher temperatures with increasing pressures. 
Such pressure dependence of the local minimum and the following upturn in $C_{11}$ seems to be related to the theoretical proposal of a pseudogap phase in URu$_2$Si$_2$ \cite{Balatsky}. On the other hand, the softening between the local maximum $\sim$120 K to the local minimum $\sim$30 K in $C_{11}$ remains even under a pressure of $\sim$1.6 GPa where the HO changes to AFM. This suggests that a lattice instability with $\Gamma_3$ symmetry could be the origin of the softening of the $C_{11}$ and $(C_{11}-C_{12})/2$ modes and persists in the region where the ground state changes from the HO to the LMAF state.
In Fig. 2(b), we represent the temperature-(hydrostatic) pressure phase diagram compiled by previous reports regarding hydrostatic measurements and present elastic anomalies found in the  $C_{11}$. The phase boundaries determined by the elastic anomaly found in the present measurements are consistent with the previous results, which were observed by using a Fluorinert mixture or Daphne oil 7373 as pressure-transmitting medium. Recently, the issue of hydrostaticity of the pressure is though discussed for the onset of the LMAF and superconducting phase in URu$_2$Si$_2$ \cite{Butch2010}. Further measurements of other ultrasonic modes under hydrostatic pressures or uniaxial pressures are now in progress to investigate the issue and also a possible ultrasonic-mode-difference under pressures in more detail.

\begin{figure}
\begin{center}
\resizebox*{13cm}{!}{\includegraphics{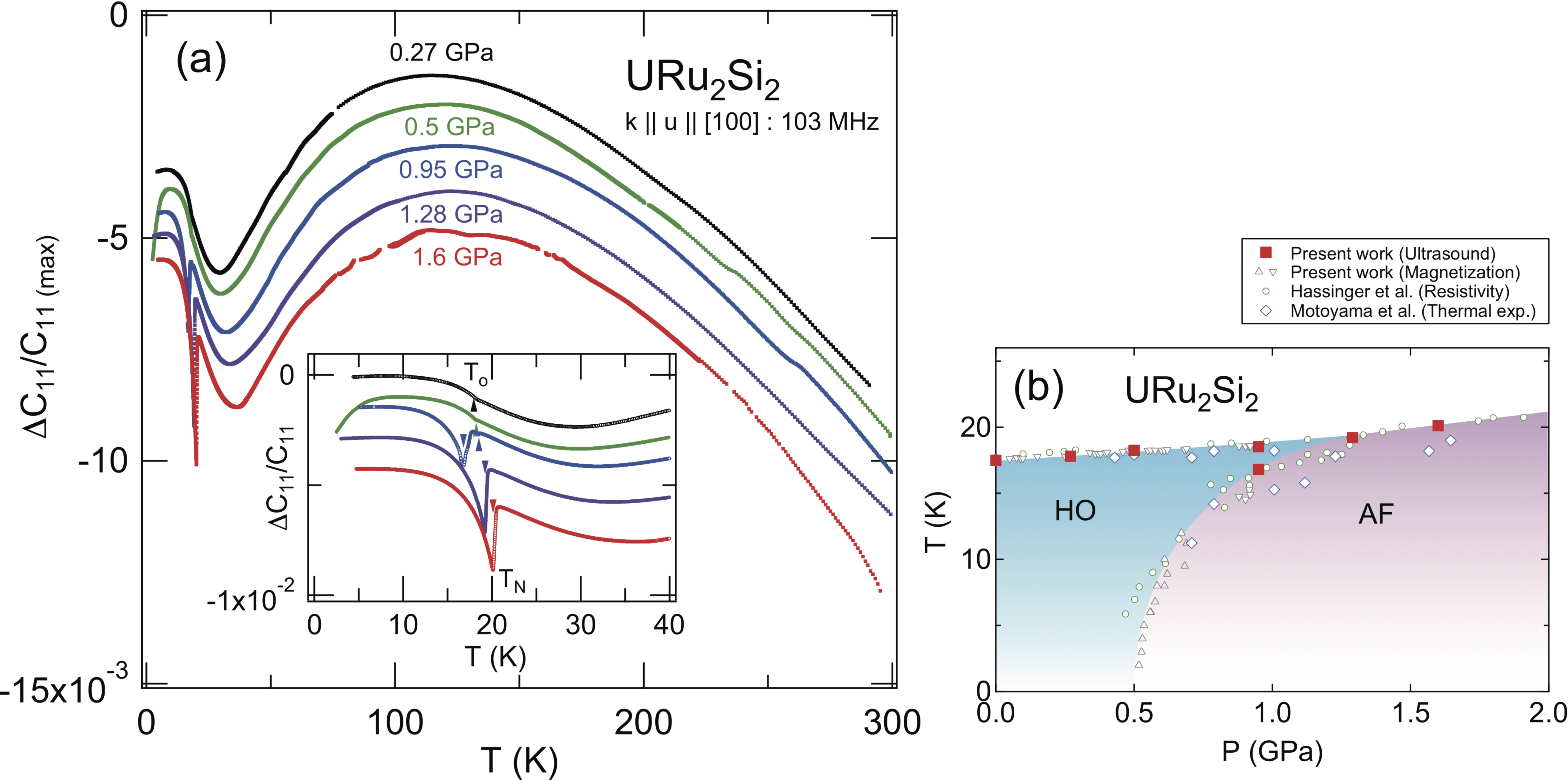}}
\caption{(a) Relative change of the elastic constant $C_{11}$ vs. temperature under several hydrostatic pressures. Inset shows zooming up of the data below 40 K. (b) Temperature-(hydrostatic) pressure phase diagram of URu$_2$Si$_2$ \cite{33, 34, 35,36}}%
\label{fig02}
\end{center}
\end{figure}

\section{Rh-doping Effect on Elastic Constants}

The Rh-doping allows us to investigate the competition of the HO and LMAF phases at ambient pressure. In the Rh-substitution system U(Ru$_{1-x}$Rh$_x$)$_2$Si$_2$, the HO phase is suppressed with increasing Rh concentration in the region of $x \le$ 0.03, while a volume fraction of LMAF phase in the HO increases with a maximum at around $x$ = 0.02 (see Fig. 3 (b) for $T$-$x$ phase diagram)\cite{37, 38}. Such drastic change from HO to LMAF in the Rh-doped samples resembles the pure URu$_2$Si$_2$ under hydrostatic pressures, as described above. Thus, we performed ultrasonic measurements on U(Ru$_{1-x}$Rh$_x$)$_2$Si$_2$ ($x$ = 0.0, 0.02, and 0.07) in order to investigate the elastic properties of the HO and the chemically-induced LMAF. 

\begin{figure}
\begin{center}
\resizebox*{13cm}{!}{\includegraphics{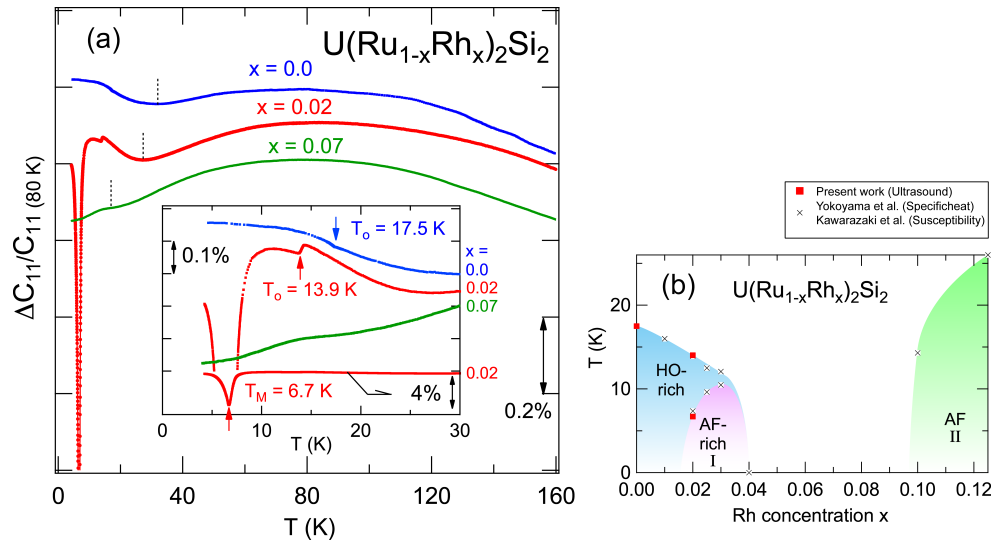}}
\caption{(a) Relative change of the elastic constant $C_{11}$ as a function of temperature in U(Ru$_{1-x}$Rh$_x$)$_2$Si$_2$ for $x$ = 0.0, 0.02 and 0. 07. Inset shows enlarged plots of the data below 30 K. (b) Temperature-Rh concentration ($x$) phase diagram of U(Ru$_{1-x}$Rh$_x$)$_2$Si$_2$\cite{37, 38,39}.}
\label{fig03}
\end{center}
\end{figure}

Figure 3 (a) shows a comparison of elastic constants $C_{11}$ for U(Ru$_{1-x}$Rh$_x$)$_2$Si$_2$ ($x$ = 0.0, 0.02, and 0.07) as a function of temperature. The data are shifted vertically for clarity. The $x$ = 0.02 sample shows a small step-like depression in $C_{11}$ at $T_{\rm o}$ = 13.9 K (as shown in the inset of Fig. 3 (a)) and a large softening of 3.2\% from 10 K to $T_{\rm M}$ = 6.7 K, the LMAF transition temperature, associated with an ultrasonic attenuation. The double transition feature of $C_{11}$ in the $x$ = 0.02 sample is similar to the pure sample ($x$ = 0.0) under 0.95 GPa (in Fig. 2 (a)), which implies that nature of the HO phase is also changed by Rh-doping [40]. It should be noted that the change at $T_{\rm M}$ for $x$ = 0.02 is one order of magnitude larger than that of $x$ = 0.0 under pressure. On the other hand, the $x$ = 0.07 does not show a local minimum but does show a broad shoulder at $\sim$20 K (indicated by the dotted line). Since both HO and LMAF are expected to be suppressed in the $x$ = 0.07 sample, the presence of a low-temperature upturn in $C_{11}$ only in the $x$ = 0.0 and 0.02 samples strongly suggests that the HO accompanies a precursor bulk modulus change. Such a volume change implies the existence of a Fermi surface instability and its reconstruction toward the HO transition. Together with the clues in the above mentioned hydrostatic pressure effects, the low-temperature upturn of $C_{11}$ seems to be connected with the pseudogap developing due to HO parameter fluctuation suggested by Haraldsen {\it et al.} \cite{Balatsky}. Based on this theory, the precursor bulk modulus change just above $T_{\rm o}$ should be tested.

\begin{figure}
\begin{center}
\resizebox*{8cm}{!}{\includegraphics{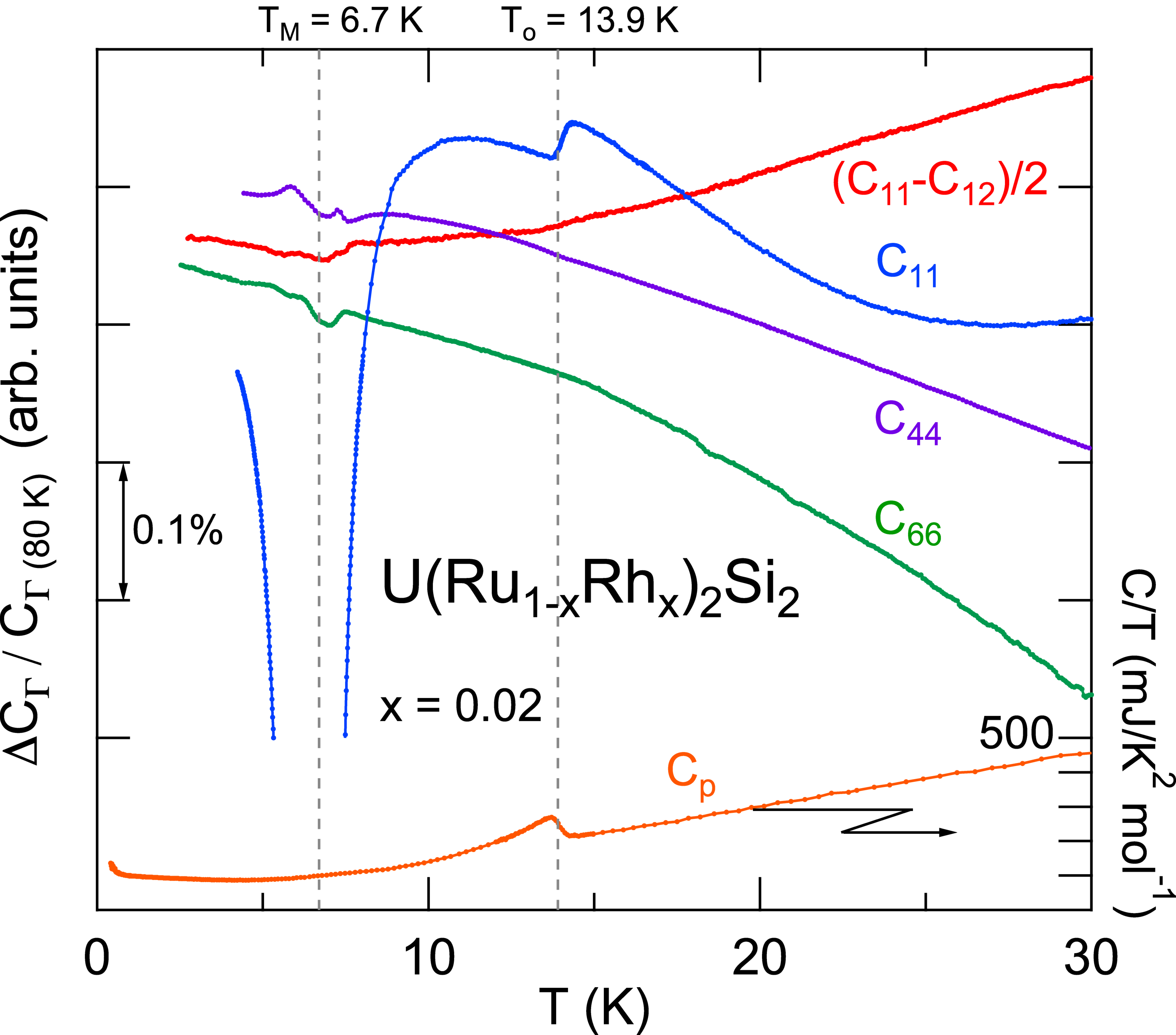}}
\caption{Relative change of the elastic constants as a function of temperature in U(Ru$_{1-x}$Rh$_x$)$_2$Si$_2$ with $x$ = 0.02. Dashed vertical lines show AFM and HO transition temperatures. Specific heat of the same Rh-concentration sample is shown (lower data as right axis) for comparison.}%
\label{fig04}
\end{center}
\end{figure}

Figure 4 shows a comparison of elastic anomalies at two successive transitions in $C_{11}$ and three shear modes $C_{44}$, $C_{66}$ and $(C_{11}-C_{12})/2$ of U(Ru$_{1-x}$Rh$_x$)$_2$Si$_2$ ($x$ = 0.02). In order to see the tiny anomalies more precisely, a higher frequency of $f \sim$ 170 MHz and a surfaced-polished single crystal was used. The dip on the elastic constant at $T_{\rm M}$ in the shear modes $(C_{11}-C_{12})/2$, $C_{44}$, and $C_{66}$ are two orders of magnitude less than the amount of the relatively large softening $\sim$3.2\% and minimum in the longitudinal $C_{11}$. Although these transverse elastic waves couple with volume conservative and symmetry breaking strains, these tiny anomalies of shear modes are understood by volume change effects due to magnetostriction in the transverse sound velocity measurement and suggest that the LMAF transition induced by Rh-doping does not accompany the symmetry breaking of the lattice, {\it i.e.}, the system seems to keep its tetragonal symmetry, within the present accuracy of the ultrasonic measurements. These results are consistent with the report of thermal-expansion measurements in the pressure-induced LMAF phase in a pure URu$_2$Si$_2$ sample under hydrostatic pressure \cite{36}. It is also indisputable that the relatively small elastic anomaly for all shear modes at $T_{\rm o}$ indicates that there is less possibility of lattice symmetry breaking in the HO phase in the 2\%-Rh doped samples.

\begin{figure}
\begin{center}
\resizebox*{8cm}{!}{\includegraphics{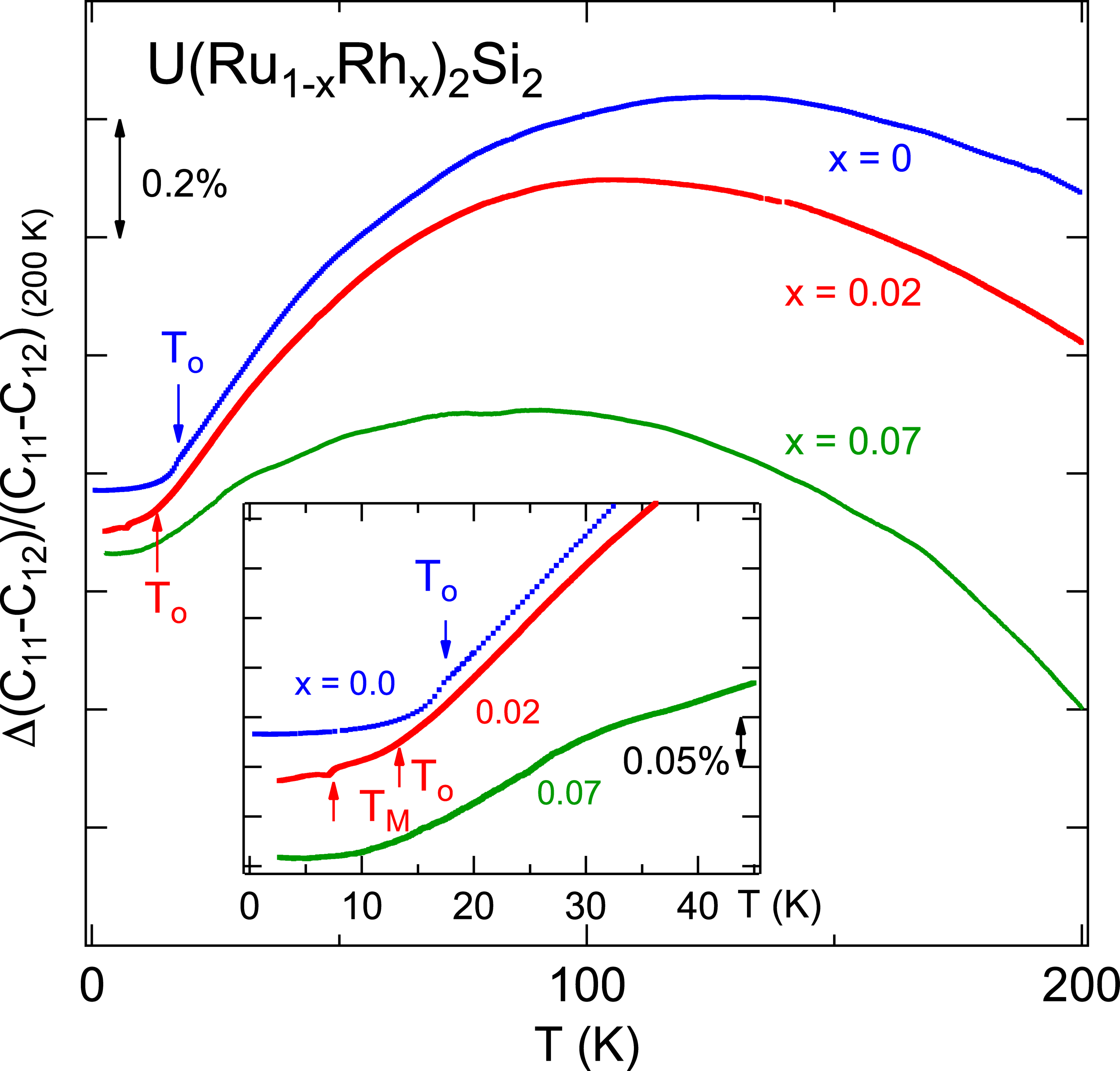}}
\caption{Relative change of the elastic constant $(C_{11}-C_{12})/2$ as a function of temperature in U(Ru$_{1-x}$Rh$_x$)$_2$Si$_2$ for $x$ = 0.0, 0.02 and 0.07. Inset shows zooming up of the data below 30 K.}%
\label{fig05}
\end{center}
\end{figure}

Next, we focus on the Rh-doping effect on the pure $\Gamma_3$-symmetry mode $(C_{11}-C_{12})/2$. Figure 5 shows $(C_{11}-C_{12})/2$ of U(Ru$_{1-x}$Rh$_x$)$_2$Si$_2$ for $x$ = 0.0, 0.02 and 0.07 as a function of temperature. The elastic anomaly at $T_{\rm o}$ in $x$ = 0.02 is visibly weaker than that of $x$ = 0.0 (see inset of Fig. 5). The amount of change in the softening above $T_{\rm o}$ is also reduced by Rh-doping, but still remains in $x$ = 0.07, where both the HO and the LMAF phases are suppressed, which is consistent with the fact that the hybridization also persists in $x$ = 0.07. Since the softening appears to be sensitive to chemical pressure and/or carrier doping such as small amounts of Rh substitution, these results advocate again that the energy gain achieved through the formation of the $c$-$f$ hybridized band leads to this Jahn-Teller type potential deformation with $\Gamma_3$ symmetry; this is a more feasible explanation for the Rh-doping effect on the softening of $(C_{11}-C_{12})/2$ than CEF effect.
A broad shoulder at $\sim$20 K in $x$ = 0.07 could be the origin of a similar elastic anomaly found in $C_{11}$. Here, it is known that another AFM phases (II or III) will appear in higher Rh concentrations (II: $0.1 \le x \le 0.3$, III: $0.4 \le x \le 1.0$) while the LMAF phase (I) is suppressed over $x \sim$ 0.04 in U(Ru$_{1-x}$Rh$_x$)$_2$Si$_2$ (see Fig. 3(b))\cite{39}.Thus, we expect that the small volume fraction of the high-Rh concentration in the sample will induce the AF phase II in $x$ = 0.07 samples, which results in the low-temperature shoulder in the elastic constants even in the shear $(C_{11}-C_{12})/2$ mode.

\section{Ultrasonics under Pulsed-Magnetic Field}

Elastic properties in high magnetic fields provide us with additional information about the relationship between the HO and the orthorhombic-symmetry lattice instability in URu$_2$Si$_2$. In magnetic fields over 35 T, along the $c$-axis of URu$_2$Si$_2$, three-successive transitions have been observed in electrical resistivity, specific heat, magnetization, thermal expansion, and ultrasonic-velocity measurements \cite{41,42,43,44}. This cascade of transitions occurs in the vicinity of the upper phase boundary of the HO phase and are associated with meta-magnetic-like increases in the $c$-axis magnetization. Magnetic field and temperature dependence of these physical properties with a wide temperature range strongly suggest that tuning URu$_2$Si$_2$ by means of a magnetic field along the $c$-axis decreases the hybridization between $5f$ and conduction electrons and leads to a reduced effective electron mass, as indicated by the disappearance of the heavy band with the collapse of the HO phase\cite{45}. Few reports concerning ultrasonic studies of the cascade transitions in the high field region using the longitudinal $C_{11}$ and $C_{33}$ modes have been reported thus far \cite{7,8,9,10,11}. But the longitudinal $C_{11}$ and $C_{33}$ modes mainly provide information about a bulk modulus change, which is complementary to thermal expansion \cite{46}. In this chapter, we review the recently performed pulsed-magnetic field measurements of the transverse-ultrasonic mode $(C_{11}-C_{12})/2$ on URu$_2$Si$_2$ for $H \parallel c$ and $H \bot c$ using long-pulsed-magnetic fields \cite{14, 47}.

\begin{figure}
\begin{center}
\resizebox*{13cm}{!}{\includegraphics{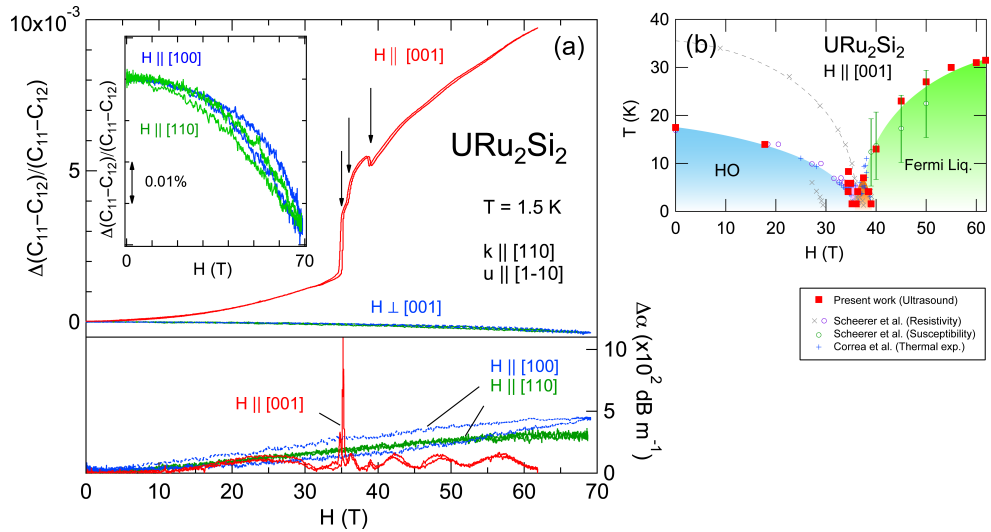}}
\caption{(a) (Upper panel) Relative change of the elastic constant $(C_{11}-C_{12})/2$ as a function of magnetic fields $H \parallel$ [001], [110] and [100]. Inset shows zooming up of the data for in $c$-plane magnetic field. (Lower panel) Ultrasonic attenuation coefficient change $\Delta \alpha$ vs. magnetic field. (b) Temperature-magnetic field phase diagram of URu$_2$Si$_2$ for $H \parallel$ [001] \cite{45,46,47}.}
\label{fig06}
\end{center}
\end{figure}

Figure 6 (a) shows the change of the elastic constant (upper panel) and ultrasonic attenuation coefficient (lower panel) vs. magnetic fields $H \parallel c$ and $H \bot c$ with parallel to [100] and [110] axes. The data shows both rising and falling processes in the magnetic field. Near the cascade transition region (with magnetic fields around 35 to 39 T), three elastic anomalies, which are indicated by arrows and correspond to the complex phase boundaries, are clearly identified. Small hysteresis of these transitions indicates that the isothermal condition was not disturbed in the sample during the pulsed-magnetic field sweep. On the other hand, no elastic anomaly has been observed for $H \bot  c$ up to 68.7 T as seen in the inset of Fig. 6(a), and we can at least conclude that the in-plane magnetic field enhances the softening of $(C_{11}-C_{12})/2$.
As apparent from Fig. 6(a), the magnetic field dependence of $(C_{11}-C_{12})/2$ for $H \parallel c$ looks very similar to the magnetization vs. $H$, and thus is reminiscent of magneto-elastic coupling. We can roughly estimate the effect as a magnetostriction on the elastic constant by using a modified-Ehrenfest relation \cite{48} written as;
\begin{equation}
        \frac{\partial H^*}{\partial p}=VT_{\rm N} \frac{\Delta \lambda_i}{\Delta [\partial M/\partial (\mu_0 H)]}.
\end{equation}
Here, $H^*$ is the phase-transition magnetic field, $p$ is hydrostatic pressure, $V$ is volume, $T_{\rm N}$ is transition temperature, $\partial M/\partial (\mu_0 H)$ is magnetic susceptibility, $\lambda = (1/L) \partial L/\partial (\mu_0 H)$ is the magnetostriction constant, estimated from the results of thermal expansion measurements under static magnetic field up to 45 T and the $c$-axis pressure dependence of the metamagnetic-transition field \cite{46, 49}. We can conclude that the influence of the magneto-striction on the elastic constant $(C_{11}-C_{12})/2$ should be negligible, {\it i.e.}, the drastic change $6 \times10^{-3}$ observed in our measurements in $(C_{11}-C_{12})/2$ toward the cascade transition region is an order of magnitude larger than the estimated change of $\sim10^{-4}$ due to magneto-striction, and therefore suggests the presence of an additional effect.
Here, we discuss one of the possible candidates for the HO-parameter, the electric antiferro-hexadecapole order model \cite{50}. This model, with an appropriate CEF level scheme, can reproduce not only the strong anisotropy of the $c$-axis magnetization but also the $T$-$P$ phase diagram and the softening of the $(C_{11}-C_{12})/2$ in zero magnetic field. It is noted that this theory also predicts that the magnetic field along the [110]-axis induces the quadrupole moment $O_2^2$ via a Ginzburg-Landau coupling term for the free energy, which should make a difference in the elastic responses, depending on the magnetic field direction. Calculation of the quadrupole susceptibility using mean-field approximation, which includes AF hexadecapole-type interactions, shows a clear difference between [100] and [110] directions, but it is still smaller than the accuracy of the present measurement \cite{14}.  While higher magnetic fields will be required to ultimately rule out the existence of hexadecapole order, this suggests that the HO is not explained by this model there is some as-yet unconsidered effect of $c$-$f$ hybridization.

\begin{figure}
\begin{center}
\resizebox*{12cm}{!}{\includegraphics{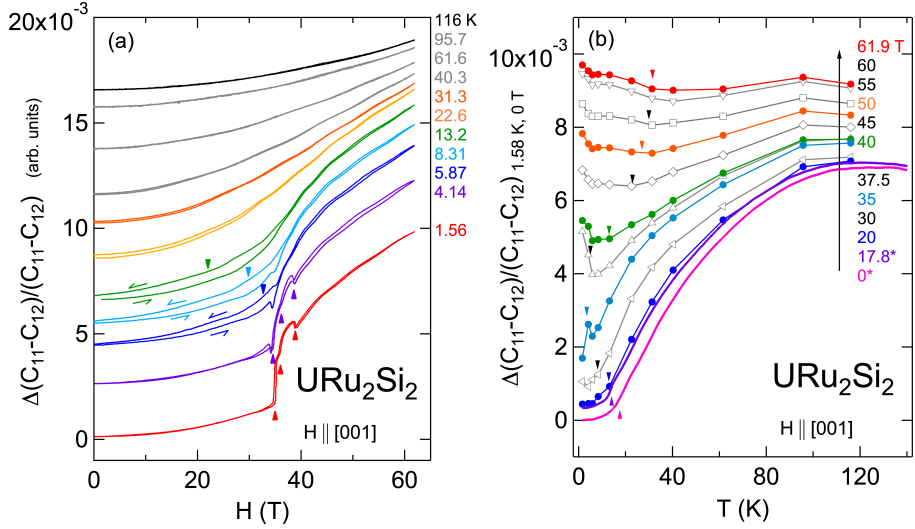}}
\caption{(a) Relative change of elastic constant $(C_{11}-C_{12})/2$ vs. the magnetic field $H \parallel$ [001] of URu$_2$Si$_2$ at various fixed temperatures between 1.56 and 116 K. The data were taken from both up- and down-sweeps of the magnetic fields, as indicated by arrows.  (b) Change of the Elastic constant vs. temperature at several fixed magnetic fields, which is converted from the data of magnetic field dependence of Fig. 7(a).}%
\label{fig07}
\end{center}
\end{figure}

Next, we explore a wider temperature range from 1.5 to 116 K, up to 61.8 T with $H \parallel$ [001] for ultrasonic measurements on URu$_2$Si$_2$, in order to check the temperature dependence of the $\Gamma_3$-lattice instability in high magnetic fields (40 $\le H \le$ 61.85 T), where both the HO and hybridization effect are suppressed. Figure 7(a) shows observed isotherms of the elastic constant $(C_{11}-C_{12})/2$ as a function of magnetic field $H \parallel$ [001] up to 61.8 T. Both curves for rising and falling magnetic fields are plotted and have been shifted vertically for clarity.
When increasing the temperature, a clear hysteresis is observed below 38 T for 5 to 13 K. These temperature and magnetic-field regions could correspond to a quantum-fluctuation region surrounding the quantum-critical end point as mentioned in Ref. \cite{51}. At higher temperature, the elastic constant increases with increasing magnetic field, but only exhibits a monotonic change. From the plot of Fig. 7(a), we can convert the data of the magnetic field dependence to a temperature dependence in various fixed magnetic fields, as shown in Fig. 7 (b). Here, the data at zero magnetic field and 17.8 T of static magnetic field are represented for comparison. The 0.7\% softening in the change of $(C_{11}-C_{12})/2$ below 120 K at zero magnetic field is gradually suppressed with increasing magnetic field. Above 40 T, $(C_{11}-C_{12})/2$ shows a minimum, which shifts to higher temperatures with increasing magnetic field.

\begin{figure}
\begin{center}
\resizebox*{8cm}{!}{\includegraphics{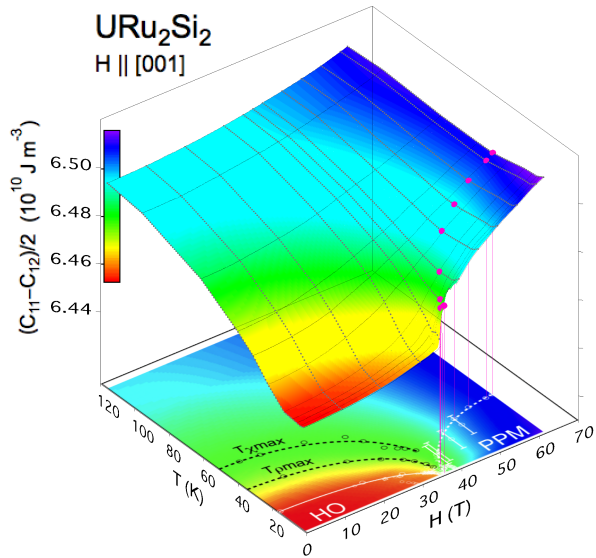}}
\caption{Three-dimensional plots of the elastic constant $(C_{11}-C_{12})/2$ vs. temperature and magnetic field along $c$ axis of URu$_2$Si$_2$. The bottom of the box shows the magnetic field-temperature phase diagram of URu$_2$Si$_2$ \cite{45,46,47}.}
\label{fig08}
\end{center}
\end{figure}

Figure 8 represents the results from Figs. 7(a) and 7(b) as three-dimensional plots. The temperature where $(C_{11}-C_{12})/2$ shows a local minimum, indicated by purple dots, increases with increasing magnetic field. The dotted curve on the basal plane represents the projection of the local minimum on the $H$-$T$ phase diagram. This curve compares favorably with the inflection point of the magnetization in the magnetic susceptibility versus $T$ curve. 
It can be clearly confirmed in Fig. 8 that the softening of $(C_{11}-C_{12})/2$ is enhanced in the red-colored region in the temperature and magnetic field region, in which the URu$_2$Si$_2$ exhibits the HO phase, where strong hybridization is also developed \cite{45}. Conversely, we can conclude that the lattice instability with $\Gamma_3$-symmetry disappears at high temperatures and high magnetic fields, where the HO phase is collapsed. The microscopic interpretation of the softening in $(C_{11}-C_{12})/2$ is still an open question and must be refocused to elucidating the nature of HO. On the other hand, the $\Gamma_3$-symmetric instability of the electron systems is not consistent with the symmetries of the recently proposed the $\Gamma_5^-$ or $\Gamma_4^-$-type HO parameters \cite{25, 52}. Further ultrasonic measurements in pulsed-magnetic fields are now in progress to compare the elastic response for different symmetry-strain fields in URu$_2$Si$_2$, which will highlight these results.

\section{Summary}

We have measured the elastic constants of URu$_2$Si$_2$ and Rh-substituted systems by means of ultrasonic measurements under hydrostatic pressure, static- and pulsed-magnetic fields up to 67.8 T over a wide temperature range. We found that an elastic softening of the $(C_{11}-C_{12})/2$ mode appears and is enhanced only in the temperature and magnetic field regions in which URu$_2$Si$_2$ exhibits a heavy-electron state. This change in $(C_{11}-C_{12})/2$ appears to be a response of the $5f$-electrons to an orthorhombic and volume conservative strain $\epsilon_{xx}-\epsilon_{yy}$ with $\Gamma_3$-symmtery. The present result advocates that the orthorhombic lattice instability is likely related to a symmetry-breaking band instability driven by $c$-$f$ hybridization and appears to be closely tied to the hidden-order of this compound.\\

The author would like to thank H. Saitoh, Y. Watanabe, S. Mombetsu (Hokkaido Univ.), who contributed to the construction of the ultrasonic apparatus, and K. Hiura and T. Murazumi (Hokkaido Univ.) for helping with ultrasonic measurements. The author also thanks W. Knafo (LNCMI, Toulouse), and H. Kusunose (Ehime Univ.) for fruitful discussions. The ultrasonic measurements and sample growth done at Hokkaido University were performed in collaboration with M. Yokoyama (Ibaraki Univ.), Y. Ikeda (ISSP, Univ. of Tokyo), H. Hidaka, and H. Amitsuka (Hokkaido Univ.). Pulsed-magnetic field measurements were performed in collaboration with M. Akatsu (Niigata Univ.), S. Yasin, S. Zherlitsyn, and J. Wosnitza (Helmholtz-Zentrum Dresden-Rossendorf, Dresden). A single crystal used in the pulsed-field measurements was grown at Univ. of California San Diego, in cooperation with M. B. Maple, K. Huang (UC San Diego) and M. Janoschek (Los Alamos Natl. Lab.). Experiments performed at Hokkaido University were supported by JSPS KAKENHI Grant No. 20740192, 19340086, 23740250 and 23102701. Experiments performed in the United States were supported by U.S. DOE. Grant No. DE-FG02-04- ER46105. We also acknowledge the support of the Hochfeld-Magnetlabor Dresden at HZDR, a member of the European Magnetic Field Laboratory.

\label{lastpage}

\end{document}